\documentclass[aps,prc,twocolumn,showpacs,superscriptaddress]{revtex4}
\usepackage{graphicx}
\usepackage{amsmath}

\begin{document}


\title{New Test of Supernova Electron Neutrino Emission using Sudbury
Neutrino Observatory Sensitivity to the Diffuse Supernova Neutrino
Background}

\author{John F. Beacom}
\email{beacom@mps.ohio-state.edu}
\affiliation{Department of Physics, The Ohio State University,
Columbus, OH 43210, USA}
\affiliation{Department of Astronomy, The Ohio State University,
Columbus, OH 43210, USA}

\author{Louis E. Strigari}
\email{strigari@mps.ohio-state.edu}
\affiliation{Department of Physics, The Ohio State University,
Columbus, OH 43210, USA}

\date{18 August 2005}

\begin{abstract}
Supernovae are rare nearby, but they are not rare in the Universe, and
all past core-collapse supernovae contributed to the Diffuse Supernova
Neutrino Background (DSNB), for which the near-term detection
prospects are very good.  The Super-Kamiokande limit on the DSNB
electron {\it antineutrino} flux, $\phi(E_\nu > 19.3 {\rm\ MeV}) <
1.2$ cm$^{-2}$ s$^{-1}$, is just above the range of recent theoretical
predictions based on the measured star formation rate history.  We
show that the Sudbury Neutrino Observatory should be able to test the
corresponding DSNB electron {\it neutrino} flux with a sensitivity as
low as $\phi(22.5 < E_\nu < 32.5 {\rm\ MeV}) \simeq 6 $ cm$^{-2}$
s$^{-1}$, improving the existing Mont Blanc limit by about three
orders of magnitude.  While conventional supernova models predict
comparable electron neutrino and antineutrino fluxes, it is often
considered that the first (and forward-directed) SN 1987A event in the
Kamiokande-II detector should be attributed to electron-neutrino
scattering with an electron, which would require a substantially
enhanced electron neutrino flux.  We show that with the required
enhancements in either the burst or thermal phase $\nu_e$ fluxes, the
DSNB electron neutrino flux would generally be detectable in the
Sudbury Neutrino Observatory.  A direct experimental test could then
resolve one of the enduring mysteries of SN 1987A: whether the first
Kamiokande-II event reveals a serious misunderstanding of supernova
physics, or was simply an unlikely statistical fluctuation.  Thus the
electron neutrino sensitivity of the Sudbury Neutrino Observatory is
an important complement to the electron antineutrino sensitivity of
Super-Kamiokande in the quest to understand the DSNB.
\end{abstract}



\maketitle


\section{Introduction} 

The core-collapse death of a massive star and the subsequent type II
optical supernova occurs at a rate of $\sim 1$ per second in the
Universe, each releasing a prodigious blast of $\sim 10^{58}$
neutrinos and antineutrinos. While a burst of many neutrino events
would be detected from a Milky Way supernova \cite{SNBurst}, the
expected occurrence rate is only $\sim 3$ per century \cite{SNRate}.
For supernova as far away as $10$ Mpc, it should be possible to
reliably detect neutrinos just one or two at a time, perhaps as often
as once per year, with present and proposed detectors
\cite{Nearby}. However, most supernovae are vastly farther, such that
the expected number of neutrino events detected per supernova is $\ll
1$. When weighted with the total supernova rate, however, the
prospects are encouraging for the detection of the {\it Diffuse
Supernova Neutrino Background} (DSNB).  The Super-Kamiokande (SK)
limit~\cite{Malek} on the flux of DSNB electron {\it antineutrinos},
$\phi(E_\nu > 19.3 {\rm\ MeV}) < 1.2$ cm$^{-2}$ s$^{-1}$, is just
above the range of recent theoretical
predictions~\cite{Fukugita,Ando2003,Strigari,Concordance}.  In
particular, the calculation based on the most recent astronomical
data predicts that the DSNB is on the verge of
detectability~\cite{Concordance}.

Searching for the DSNB signal requires the detection of an excess
event rate over backgrounds, exploiting the different energy spectra
of the signal and backgrounds. Due to the near-isotropy of the
scattered positrons in the reaction $\bar{\nu}_e + p \rightarrow e^+ +
n$ and the large distances to the supernovae, the detections of DSNB
events will not be associated with particular optical bursts. If
Super-Kamiokande is enhanced by the addition of gadolinium, as
proposed by Beacom and Vagins, the detector backgrounds would be
greatly reduced, allowing the clean detection of as many as 6 events
per year \cite{gadzooks, Vagins}.  Detection of the DSNB may be the
first detection of neutrinos from beyond 1 Mpc, and the second
detection of supernova neutrinos.  In just a few years, the yield from
SN 1987A could be surpassed.  This will provide valuable insight into
supernova physics, neutrino properties, and the history of
cosmological massive star formation.

In the following, we discuss the prospects for detection of flavors of
the DSNB other than $\bar{\nu}_e$, which is the dominant yield in SK.
Specifically, we propose using the Sudbury Neutrino Observatory (SNO)
to study the DSNB electron {\it neutrino} flux.  SNO has a unique
sensitivity to the electron neutrino flux though the charged-current
interaction with deuterons, $\nu_e +d \rightarrow e^- + p + p$. By
combining this charged-current reaction with neutrino-electron elastic
scattering and the neutral-current breakup reaction $\nu + d
\rightarrow \nu + n + p$, SNO has studied in detail the $^8$B solar
neutrino spectrum~\cite{SNOB8}, with analysis of the $hep$ spectrum
forthcoming.  Using the theoretical $hep$ spectrum and the measured
atmospheric backgrounds above 20 MeV at SK~\cite{Malek}, we
estimate the DSNB backgrounds at SNO.  We then show that SNO could
very likely reach a sensitivity as low as $\phi( 22.5 < E_{\nu} < 32.5
\, {\rm MeV}) \simeq 6 \, {\rm cm}^{-2} \, {\rm s}^{-1}$.  This would
improve the existing limit from Mont Blanc by about three orders of
magnitude~\cite{Aglietta}, and provide an important complement to the
SK $\bar{\nu}_e$ limit. In addition to the importance of studying the
$\nu_e$ emission for supernova models, it is also crucial for testing
neutrino mixing, as stressed by Lunardini and Smirnov
\cite{Lunardini}.

There is an additional motivation to test the DSNB $\nu_e$ flux
separately from $\bar{\nu}_e$, stemming from the observed angular
distribution of the SN 1987A data. Two well-studied features of this
data that still stand out are that the first event in the
Kamiokande-II detector was scattered forward, and the time-integrated
angular distributions in both the Kamiokande-II and IMB detectors were
both generally more forward and less isotropic than expected from
$\bar{\nu}_e + p \rightarrow e^+ + n$ events alone.  Both features
have typically been explained as either an increase in the
neutrino-electron scattering rate beyond theoretical predictions, or a
statistical fluctuation~\cite{Raffelt}. With the SN 1987A data alone,
this debate cannot be resolved.

Increasing the $\nu_e$ flux to the extent that these features of the
SN 1987A data were a probable outcome is in conflict with standard
models of supernova neutrino emission.  However, given the fact that
the models generally fail to produce successful explosions, the
possibility of a serious misunderstanding of supernovae, perhaps due
to new physics, cannot be excluded.  Here we point out that such an
increase is testable, in that it would generally lead to the DSNB
$\nu_e$ flux being large enough for SNO to detect. The absence of a
signal would greatly strengthen the case that these features of the
angular distribution of the SN 1987A data were due to a statistical
fluctuation.  Resolution of this point is also important for the
future interpretation of the SK DSNB $\bar{\nu}_e$ results.

In Section \ref{sec:DSNBflux}, we review the DSNB predictions and
limits; in Section \ref{sec:SNO}, we calculate the capabilities of SNO
to detect the DSNB $\nu_e$ flux; in Section \ref{sec:1987A}, we apply
these results to discuss the future constraints on the electron
neutrino emission from SN1987A, and then we finish with our
conclusions.


\section{Present Perspective on the DSNB}
\label{sec:DSNBflux} 


\subsection{Emission per Supernova}

Stars greater than $\sim 8 M_\odot$ are able to burn elements by
successive nuclear fusion reactions until iron is reached, beyond
which point no further energy generation is possible \cite{Heger}.
Once the core has formed about a Chandrasekhar mass of iron, not even
electron degeneracy pressure can support it under the weight of the
stellar envelope, and it collapses until it reaches the density of
nuclear matter, bounces, and forms an outgoing shock.  If energetic
enough, the shock will eject the envelope and cause the optical
supernova, leaving behind a neutron star; if not, the core and
envelope will collapse into a black hole.  In either case, the $3
\times 10^{53}$ erg of gravitational binding energy release is
dominantly radiated away by all flavors of neutrinos and antineutrinos
over about 10 s (if black hole formation occurs sooner than this, then
the neutrino emission is less).  It is usually assumed that this total
energy release $E^{\rm tot}_\nu$ is shared more or less equally among
the six flavors of neutrinos and antineutrinos.  For recent updates
and reviews on supernova neutrino emission, see
Refs.~\cite{Heger,SNnu}.  The spectroscopically classified type II,
Ib, and Ic supernovae (hereafter SNII) result from massive star core
collapse, and lead to the emission of $\sim 10^{58}$ neutrinos and
antineutrinos.  Type Ia supernovae (hereafter SNIa) result from the
sudden ignition of a white dwarf accreting material from a companion,
and are not accompanied by comparable neutrino emission.

In core-collapse supernovae, neutrinos must diffuse out of the
proto-neutron star, so that they should decouple with nearly thermal
spectra, characteristic of their surface of last scattering.  Since
$\nu_\mu$ and $\nu_\tau$ and their antiparticles only have
neutral-current interactions at these low energies, they should
decouple at the smallest radius and largest temperature.  Noting the
charged-current interactions of $\nu_e$ and $\bar{\nu}_e$, and that
the medium is neutron-rich, one conventionally expected a hierarchy of
temperatures like $T_{\nu_e} \simeq 4$ MeV, $T_{\bar{\nu}_e} \simeq 5$
MeV, and ${\rm T}_{\nu_\mu} \simeq 8$ MeV.  More recent work indicates
both lower expected temperatures, as well as lower differences between
flavors~\cite{SNnu}.  Assuming equipartition of the total energy among
the six flavors, the spectrum per flavor is approximately
\begin{equation}
\frac{dN}{dE}(E) = \frac{E^{\rm tot}_\nu}{6} \, \frac{120}{7\pi^4} \,
\frac{E^2}{T^4} \, \left[\exp\left(\frac{E} {T}\right)+ 1\right]^{-1}\,.
\label{eq:spectrum}
\end{equation}
Neutrino oscillations may mix the spectra of different flavors, and
importantly, can increase the average energy of the more easily
detectable $\bar{\nu}_e$ and $\nu_e$ (for example, see
Ref.~\cite{SNosc}), and we return to this point below.


\subsection{Integration over Past Supernovae}

It has been known for some time that the star formation rate was
larger in the past, and in particular, was about an order of magnitude
larger at redshift $z \simeq1$ than it is today, likely increasing
more slowly at larger redshifts, and first turning on at an uncertain
redshift $z \agt 5$ (these data have been comprehensively reviewed by
Hopkins~\cite{Hopkins}).  In the past year, these results have been
markedly improved by new data, especially from the GALEX ultraviolet
satellite~\cite{Schiminovich:2004km}.  These and other recent
data~\cite{Baldry,Perez-Gonzalez,Bell}, while confirming the basic
picture above, also indicate that the absolute rates are on the high
side of past estimates, due to larger (and better understood)
corrections for obscuration by dust.

Interestingly, this recent trend in astronomical analyses to favor
larger dust corrections and larger star formation rates can be very
usefully constrained by the upper limits on the DSNB.  In
Ref.~\cite{Concordance}, Strigari {\it et al.} developed a
``Concordance Model" for the star formation and supernova rates which
accounts for these and other data.  This model is characterized by the
star formation rate per comoving cubic Mpc today, $R_{\rm SFR}(z = 0)
= 0.02$ Mpc$^{-3}$ yr$^{-1}$, and the growth rate $R_{\rm
SFR}(z)/R_{\rm SFR}(z = 0) = (1 + z)^\beta$, with $\beta = 2$, assumed
to hold to $z = 1$, after which a constant rate was assumed.  These
choices of $R_{\rm SFR}(z = 0)$ and $\beta$ are slightly conservative,
relative to the GALEX data (see Fig.~1 of Ref.~\cite{Concordance}).
For larger $z$, the neutrinos are typically redshifted below the
detector threshold, and so have relatively little effect on the
result.  Taking into account the stellar initial mass function, this
converts to a core-collapse rate of $R_{\rm SNII}(z = 0) = 2.5 \times
10^{-4}$ Mpc$^{-3}$ yr$^{-1}$, with the same functional form, since
SNII stellar lifetimes are short~\cite{Concordance}.  An important
test of the Concordance Model is that it agrees with measurements of
the SNII rate (and the SNIa rate too), confirming the choices of
conversion factors, and requiring that the fraction of failed
supernovae directly forming black holes~\cite{BH} must be relatively
small~\cite{Concordance}.  The neutrino emission parameters assumed
were $3 \times 10^{53}$ erg, equipartitioned among flavors, and
normal-hierarchy mixing of $T = 5$ MeV $\bar{\nu}_e$ with $T = 8$ MeV
$\bar{\nu}_\mu$ and $\bar{\nu}_\tau$.  For smaller temperatures and/or
differences between temperatures, the predicted flux would be somewhat
less.

\begin{figure}
\includegraphics[width=3.25in]{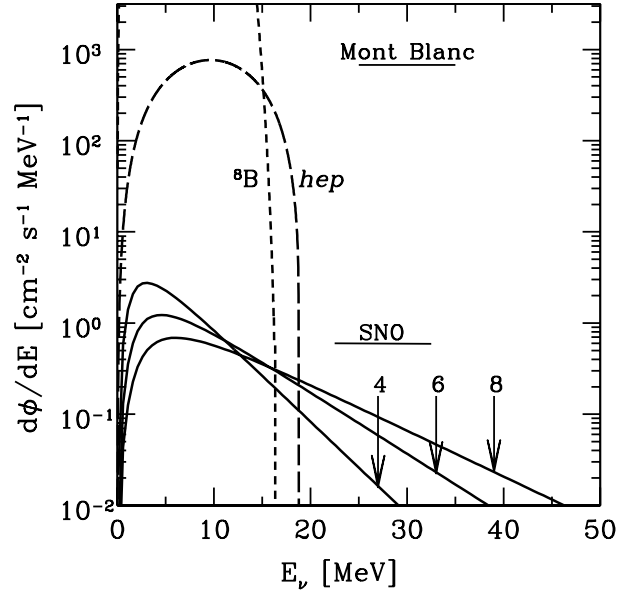}
\caption{\label{fig:Fluxes}
Predictions for the DSNB $\nu_e$ spectra assuming $T = 4, 6$, or 8
MeV (solid curves), along with the approximate published Mont
Blanc limit and the projected SNO sensitivity (solid lines).  The
solar neutrino $\nu_e$ spectra, which are irreducible backgrounds for
this study, are also shown.  The atmospheric neutrino fluxes are not
shown here, but their effects on the detectable spectra are shown in
Fig.~\ref{fig:Rates}.}
\end{figure}

Using the Concordance Model~\cite{Concordance} results for the SNII
rate convolved with the neutrino emission per supernova $dN/dE$, the
expected DSNB differential flux $d\phi/dE$ is
\begin{equation}
\frac{d\phi(E)}{dE} = \int \! R_{\rm SNII}(z)
\frac{dN(E(1+z))}{dE} (1+z) \frac{dt}{dz} dz\,,
\label{eq:flux}
\end{equation}
where the integral runs from redshift zero to at least $z \simeq 1$,
beyond which there is little contribution.  To determine $dt/dz$, we
use a $\Lambda$-CDM cosmology, $\Omega_{{\rm M}} = 1- \Omega_{\Lambda} =
0.3$, and $H_0 = 70$ km s$^{-1}$ Mpc$^{-3}$, so that
\begin{equation} 
\left \vert \frac{dt}{dz} \right \vert = \frac{1}
{{\rm H}_{0} (1+z) \sqrt{ \Omega_{{\rm M}} (1+z)^{3} + \Omega_{\Lambda}}}\,.
\label{eq:lcdm}
\end{equation}
Our results are in good general agreement with other recent
calculations up to different choices of the
inputs~\cite{Fukugita,Ando2003,Strigari}; these results and others are
nicely reviewed by Ando and Sato~\cite{AndoReview}.  The key input is
the star formation rate at redshift zero, and the most recent
astronomical measurements used in Ref.~\cite{Concordance} are both
larger and more precise than those used in
Refs.~\cite{Fukugita,Ando2003,Strigari,AndoReview}.  In
Fig.~\ref{fig:Fluxes}, we show predictions for the DSNB $\nu_e$
fluxes, for a few possible choices of the $\nu_e$ temperature, without
including any specific neutrino oscillation scenario yet.


\subsection{Existing Limits on the DSNB}

The SK upper limit on the DSNB flux of $\bar{\nu}_e$ is by far the
most stringent for any flavor, and is $1.2$ cm$^{-2}$ s$^{-1}$ for
energies above 19.3 MeV~\cite{Malek}.  (With increasing energy, the
signal falls and the background rises, and so the flux at energies
beyond about 30 MeV is less important.)  The detection channel in SK
is inverse beta decay on free proton targets, $\bar{\nu}_e + p
\rightarrow e^+ + n$, and in the existing analysis, only the positron
is detected~\cite{Malek}.  At the present level of uncertainty and low
statistics, the recoil-order corrections to the cross section and
kinematics~\cite{invbeta} can be ignored, so that $E_e \simeq E_\nu -
1.3$ MeV, where $E_e$ is the positron total energy, following the SK
convention.  Other singles rates in the same energy range as the
signal create serious backgrounds.  Below $E_e = 18$ MeV, uncut
muon-induced beta radioactivities and solar neutrino events overwhelm
the DSNB signal.  At higher energies, the dominant background is from
the decays of sub-\v{C}erenkov muons, produced by atmospheric
$\nu_\mu$ and $\bar{\nu}_\mu$ interacting inside the fiducial volume.
These non-relativistic muons quickly lose energy and decay at rest,
with the relativistic electrons and positrons following the well-known
Michel spectrum.  Since their spectrum is known, and the total rate
can be normalized from the data, the results are nearly independent of
atmospheric neutrino flux uncertainties.  The atmospheric $\nu_e$ and
$\bar{\nu}_e$ fluxes create a smaller background that can be well fit
at energies above the Michel peak.  Figure~2 of Ref.~\cite{Malek} is a
clear illustration of these backgrounds and their large impact on the
DSNB signal sensitivity.  Importantly, recent DSNB
calculations~\cite{Fukugita,Ando2003,Strigari,Concordance,AndoReview}
are not far below the SK limit, and the Concordance Model result based
on the latest astronomical data is especially
close~\cite{Concordance}.

The prospects for the future may be even brighter.  Beacom and Vagins
proposed that if SK were enhanced by the addition of dissolved
gadolinium trichloride (GdCl$_3$), then it would be possible to tag
the neutrons produced in the signal reaction, greatly reducing the
background rates~\cite{gadzooks}.  Gadolinium has a huge cross section
for radiative neutron capture, and the produced gamma rays would
Compton scatter electrons, producing detectable \v{C}erenkov light.
With much-reduced backgrounds, a more favorable energy range could be
used (beginning near 10 MeV), and SK could cleanly detect as many as 6
DSNB $\bar{\nu}_e$ events per year; see Fig.~1 of
Ref.~\cite{gadzooks}.  This could lead to the first detection of DSNB
neutrinos.  While gadolinium compounds have been commonly used in
oil-based detectors, Ref.~\cite{gadzooks} was the first to propose
that dissolved gadolinium could be added to a water-based detector,
and that this is the only cost-feasible option to develop a
neutron-sensitive detector as large as SK, or ultimately at the 1-Mton
scale~\cite{gadzooks}.  The research and development work continues to
be very encouraging, supporting the claims that the required standards
of solubility, ease of use, water transparency, radiopurity, effects
on detector materials, cost, and safety will be
met~\cite{gadzooks,Vagins}.  As a full system test, in late 2005
Vagins will add GdCl$_3$ to the 1-kton water-\v{C}erenkov detector at
KEK (a scale model of SK, and a former near detector of the K2K
long-baseline neutrino experiment~\cite{K2K}).

The only published limits on the DSNB fluxes of other flavors come
from the Mont Blanc experiment~\cite{Aglietta}.  Their limit on
$\nu_e$ flux is $\phi(25 < E_\nu < 50 {\rm\ MeV}) < 6.8 \times 10^3$
cm$^{-2}$ s$^{-1}$, and is based on $\nu_e$ charged-current
interactions with $^{12}$C.  Since the DSNB spectra are quickly
falling, even when weighted with the detection cross section, nearly
all of their sensitivity would have come from the beginning of the
above energy range.  To make a more direct comparison with the
sensitivity we derive for SNO, we simply assume that the Mont Blanc
limit also applies to an interval of width 10 MeV, i.e., from 25 to 35
MeV.  In Fig.~\ref{fig:Fluxes}, we show both the Mont Blanc limit and
the projected SNO sensitivity as if they were constant over 10 MeV,
which is an approximation (e.g., note Fig.~\ref{fig:Rates}).

There is also a Mont Blanc limit on the $\nu_\mu$ and $\nu_\tau$ flux,
based on neutral-current interactions with $^{12}$C, and this is
$\phi(20 < E_\nu < 100 {\rm\ MeV}) < 3 \times 10^7$ cm$^{-2}$ s$^{-1}$,
with a similar limit for $\bar{\nu}_\mu$ and $\bar{\nu}_\tau$.  These
are very weak limits (for comparison, the solar $^8$B flux is $5
\times 10^6$ cm$^{-2}$ s$^{-1}$), and could be greatly improved.  For
example, a DSNB flux as large as the Mont Blanc limit would cause a
huge excess of neutral-current deuteron breakup events in SNO, at
least $10^3$ neutrons {\it per day}.  It may be sufficient to set
limits on the fluxes of DSNB $\bar{\nu}_e$ and $\nu_e$ and use
knowledge of the neutrino mixing angles to deduce limits on the DSNB
fluxes of $\nu_\mu$, $\nu_\tau$, and their antiparticles.


\section{DSNB Detection in SNO}
\label{sec:SNO} 

  
\subsection{Detection of $\nu_e + d \rightarrow e^- + p + p$}

SNO, the first water-\v{C}erenkov detector to use deuterons as a
target for astrophysical neutrinos, is based on 1 kton of heavy water,
D$_2$O~\cite{SNO,Aharmim2005}.  The deuterons are targets for the
charged-current reactions $\nu_e + d \rightarrow e^- + p + p$ and
$\bar{\nu}_e + d \rightarrow e^+ + n + n$; electrons and positrons are
detected by their \v{C}erenkov light, as in light-water detectors like
SK.  Additionally, SNO has the ability to detect neutrons, so that it
can measure the neutral-current reactions $\nu + d \rightarrow \nu + p
+ n$ and $\bar{\nu} + d \rightarrow \bar{\nu} + p + n$, sensitive to
all flavors of neutrinos and antineutrinos, and separate them from the
two charged-current reactions.  Three neutron detection techniques
have been used in SNO: radiative neutron capture on deuterons in the
pure D$_2$O phase, radiative neutron capture on chlorine in the D$_2$O
plus dissolved NaCl phase, and capture on discrete $^3$He-based
neutron counters at present~\cite{SNO,Aharmim2005}.  In all cases,
neutrons are simply counted, with no energy or direction information
available.

In a search for $\nu_e$, SNO has the advantage of the large and
spectral (good fidelity between incoming neutrino and electron energy)
cross section on deuterons; SK depends on the much less favorable
neutrino-electron scattering cross section.  In contrast, in a search
for $\bar{\nu}_e$, SK has the advantage of a much larger detector
size, as well as a more favorable cross section on free protons.  The
other key features of SNO are the existence of a good neutral-current
detection channel and the ability to detect neutrons.  While the
former is unique to D$_2$O, SK may soon have the latter
ability~\cite{gadzooks,Vagins}, leveraged by its much greater fiducial
mass of 22.5 kton.

Other authors have considered the SNO sensitivity to the DSNB
$\bar{\nu}_e$, noting that the very clean signal coincidence of a
positron and two separate neutrons should allow better background
rejection than in SK~\cite{Kaplinghat,Ando2002}.  With limited
exposure and low neutron detection efficiency (in the pure D$_2$O
phase), the SNO results so far are not very restrictive, but
significant improvements are expected~\cite{Aharmim2004}.

What we are proposing here, which has not been noted before, is that
SNO should be able to make a unique contribution by exploiting its
sensitivity to DSNB $\nu_e$.  Since this will be just a singles search
(only the electron in the final state is detectable), consideration of
backgrounds will be crucial.

Recent calculations of the neutrino-deuteron cross sections are given
in Refs.~\cite{Nakamura,Butler}; in this energy range, the $\nu_e$
cross section is $\simeq 65\%$ larger than for $\bar{\nu}_e$, and a
factor $\simeq 2$ less than the inverse beta cross section on free
protons.  The threshold for the $\nu_e + d$ interaction is 1.4 MeV,
less than the 4.0 MeV for $\bar{\nu}_e + d$ or 1.8 MeV for
$\bar{\nu}_e + p$.  In the $\nu_e$ channel, while the maximum electron
(total = kinetic plus mass) energy is $E_e = E_\nu - 0.9$ MeV, the
peak in the differential cross section is
lower~\cite{Nakamura,Butler}, and so we use $E_e = E_\nu - 2.5$ MeV to
assign energies.  For the present purposes, a delta function with
these kinematics is an adequate approximation to the full differential
cross section in the relevant energy range, and makes interpretation
of the results more straightforward.  The total event rate is
\begin{equation}
R_{\nu} = N_d  T \int 
\frac{d\phi(E_\nu)}{dE_\nu}
\frac{d\sigma(E_\nu,E_e)}{dE_e} dE_\nu dE_e\,,
\label{eq:rate} 
\end{equation}
where $N_d = 6 \times 10^{31}$ is the number of deuterons per kton of
D$_2$O, and $T$ is the exposure time.  In making our estimate of the
SNO DSNB sensitivity, we assume that the total exposure will be 5
kton-yr at full efficiency.  We emphasize that in order to more
accurately determine the true sensitivity, a comprehensive study by
the SNO collaboration is needed.

In Fig.~\ref{fig:Rates}, we show our estimates of the DSNB $\nu_e$
detection spectra for three chosen temperatures.  SNO is less
sensitive than SK, and thus can only detect the DSNB if the flux is
larger than expected.  We have thus renormalized our standard
predictions according to our detection criterion, justified in more
detail below, of 1.6 expected signal events in 5 years with detected
energy $E_e$ between 20 and 30 MeV.  A detailed discussion of the
backgrounds is needed, and we turn to that next.


\subsection{Detector Backgrounds}
\label{sec:bkgds}

\begin{figure}
\includegraphics[width=3.25in]{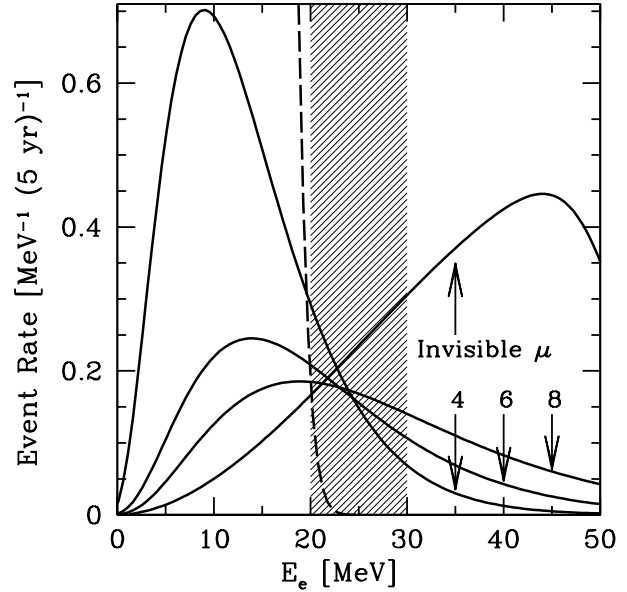}
\caption{\label{fig:Rates}
The DSNB $\nu_e$ signal spectra in SNO for a 5 kton-yr exposure, each
renormalized to give an expectation of 1.6 signal events in the
electron energy range of 20--30 MeV (shaded).  The renormalization
factors for each temperature are given in Table~\ref{tab:table1}.
Below 20 MeV, solar neutrino interactions (the $hep$ flux is shown
with the dashed line) and other backgrounds are overwhelming.  Above
30 MeV, the sub-\v{C}erenkov (invisible) muon and other atmospheric
neutrino backgrounds are too large.  The energy resolution of SNO has
been taken into account.  In the irrelevant region below 20 MeV, the
signal curves are subject to some approximations in the
calculation, and should not be taken too literally.}
\end{figure}

In Fig.~\ref{fig:Rates}, we show the renormalized DSNB $\nu_e$
spectra, along with our estimates of the most important background
processes: solar neutrinos and sub-\v{C}erenkov muons produced by
atmospheric neutrinos.  As we will show, the possible SNO sensitivity
to DSNB $\nu_e$ is several times worse than the existing SK limit on
DSNB $\bar{\nu}_e$; therefore, we ignore DSNB $\bar{\nu}_e$
interactions as a possible background.

Since the proposed detection channel is $\nu_e + d \rightarrow e^- + p
+ p$, below about 20 MeV solar neutrinos provide an overwhelming
background, even taking into account the fact that neutrino mixing
reduces the flux of $\nu_e$.  The solar neutrino fluxes~\cite{SSM} are
shown in Fig.~\ref{fig:Fluxes}.  At the highest energies, the dominant
background is from the solar $hep$ reaction ($^3{\rm He} + p
\rightarrow \, ^4{\rm He} + e^+ + \nu_e$)~\cite{hep}; even though its
flux is $\sim 10^3$ times smaller than the solar $^8$B beta
decay~\cite{8boron} flux, its endpoint is about 19 MeV, about 3 MeV
higher than for $^8$B.  Since we only consider energies above 20 MeV,
this remains true even when energy resolution~\cite{SNOB8} is taken
into account, as we do.  There is no possibility of significantly
reducing this background, since it is the same detection reaction as
the signal, and the angular distribution is only weakly backward (see
Fig.~3 of Ref.~\cite{invbeta}).  Therefore we neglect consideration of
all other backgrounds, and the precise details of the signals, below
20 MeV.  Additionally, we neglect the diffuse $\nu_e$ flux made by all
of the stars in the Universe {\it before} they end their lives (some
as supernovae), since this is buried by the solar $\nu_e$
flux~\cite{stellarnue}.

In the SK DSNB $\bar{\nu}_e$ search, the most significant background
at high energies is from the electrons and positrons produced by
sub-\v{C}erenkov muons decaying at rest.  These muons are invisible
since they are produced inside the detector fiducial volume with
non-relativistic initial energies, and they quickly stop; their decay
spectrum is simply the well-known Michel spectrum.  Since SK has
measured the rate of these decays, and their spectrum is fixed, this
greatly reduces the dependence on atmospheric neutrino flux
predictions.  The invisible muons are produced by low-energy
atmospheric $\nu_\mu$ and $\bar{\nu}_\mu$, primarily interacting with
bound neutrons and protons in oxygen nuclei.  After correcting for
detector efficiency, the measured rate in SK is about 60 events per
year inside the 22.5 kton fiducial volume~\cite{Malek}.  Taking into
account vacuum neutrino oscillations of $\nu_\mu$ into $\nu_\tau$ (and
their antiparticles), this is in good agreement with
expectations~\cite{Malek}.  There are also background events due to
very low-energy atmospheric $\nu_e$ and $\bar{\nu}_e$, also primarily
interacting with bound nucleons.  Since their rate is much smaller
than the rate due to decaying invisible muons, especially below 40
MeV~\cite{Malek}, we do not consider them further.

In the energy range 20--30 MeV, the efficiency-corrected rate of
invisible muon decays in SK is about 0.5/kton/yr.  To estimate the
rate of invisible muon decays in SNO, we have to correct for the fact
that SK and SNO are at different latitudes.  In the relevant neutrino
energy range, near 200 MeV, Ref.~\cite{Naumov} suggests a correction
of 1.8, while Ref.~\cite{BGS} suggests 1.5.  To be conservative, we
estimate that the rate will be a factor of 2 {\it larger} at the
location of SNO than in SK.  We assume that events with a final-state
neutron can be tagged, and that about half of the events produced on
oxygen can be tagged by a nuclear de-excitation gamma from the
residual $^{15}$O or $^{15}$N \cite{Kolbe}; together these lead to a
reduction factor in the background rate of about 0.5, canceling the
assumed latitude correction.  In practice, the invisible muon decay
rate will be measured and the assumed final-state probabilities
checked.  Our estimated residual invisible muon decay background rate
in SNO of 0.5/kton/yr in the range 20--30 MeV is conservative, and
will likely be reduced by several corrections we have neglected, each
at the $\sim 10-20\%$ level and favorable.  These include: a smaller
latitude correction would be more accurate, the number of oxygen
nuclei per kton of heavy water is less than in light water, $\mu^-$
capture on oxygen will reduce the number of background invisible muon
decays, and the neutrino-deuteron cross sections are less than for
free nucleons.  Also, the observing period in SNO is mostly after that
used in the SK analysis~\cite{Malek}, and closer to solar maximum,
when atmospheric neutrino fluxes are smaller, particularly at high
latitudes and low energies~\cite{atmospheric}.
 

\subsection{SNO Sensitivity to the Electron Neutrino Flux} 
\label{sec:rate}

\begin{table}
\caption{\label{tab:table1}
For each assumed temperature, we list the total DSNB flux and the flux
just in the energy range $22.5-32.5$ MeV, both in units of cm$^{-2}$
s$^{-1}$.  The last column gives the factor by which the flux would
need to be increased to meet our detection criteria.}
\begin{ruledtabular}
\begin{tabular}{ccccc}
${\rm T}$ & $\phi$ (all $E_\nu$) & $\phi$ ($22.5-32.5$ MeV) &
$\phi_{\rm sens.}/\phi_{\rm pred.}$\\
\hline
4 & 20 & 0.2 & 32\\
5 & 16 & 0.4 & 15\\
6 & 14 & 0.6 & 9\\
7 & 12 & 0.8 & 7\\
8 & 10 & 0.9 & 6\\
\end{tabular}
\end{ruledtabular}
\end{table}

To estimate the SNO sensitivity to DSNB $\nu_e$ in the energy interval
20--30 MeV, we consider two primary requirements.  First, that in the
assumed 5 kton-yr exposure, there should be at least one signal event.
Second, that the number of signal events $N_S$ should be larger than
the square root of the number of background events $N_B$ for the same
exposure (after our assumed cuts).  Since we estimate $N_B = 2.5$, the
second requirement $N_S > \sqrt{N_B} = 1.6$ is hardly different from
the first that $N_S > 1$.  Our treatment is crude, but it is not yet
possible to be more precise.  The SNO backgrounds rates, which we have
tried to estimate conservatively (see above), have not yet been
published.  Further, since both $N_S, N_B \sim 1$, the statistical
fluctuations in each will also be $\sim 1$, and what numbers of events
happen to occur will affect the final conclusions.  In addition, the
energies at which they occur will matter -- a single event at 21 MeV
would be interpreted differently than a single event at 29 MeV.  Since
the statistical errors are so large, we can neglect all other
uncertainties.  In trying to motivate a complete and sophisticated
analysis of the SNO data, we are simply trying to establish the likely
scale of the sensitivity and show why this would be an interesting and
important result.  The full analysis will have to be done by the SNO
Collaboration.

Since the sensitivity is above the level of standard predictions, we
consider how much larger the flux would have to be to yield a
detection.  This factor must be at least
\begin{equation}
\frac{\phi_{\rm sens.}}{\phi_{\rm pred.}}
\simeq \frac{\max(1,\sqrt{N_B})}{N_S}\,,
\label{eq:limit} 
\end{equation}
where inside this equation, $N_S$ is the number of signal events
predicted in the standard case.  Since $N_S, N_B$ are so small, we
neglect the possible spectral differences between signal and
background in the range 20--30 MeV, and only consider the counts, so
that the required $N_S$ does not depend on the assumed supernova
neutrino temperature.  Since the predicted flux in this range does
depend on temperature, so does the required renormalization in order
to have a detection.  These results are given in
Table~\ref{tab:table1}, as a function of the assumed temperature.  For
low temperatures, large fluxes would be required.  On the other hand,
for larger temperatures, the SNO sensitivity is reasonably close to
standard predictions.  We do not consider neutrino mixing for three
reasons.  First, the low statistics.  Second, the fact that over this
narrow energy interval, a composite spectrum could be considered to be
dominated by a largest temperature contributing to the spectrum.
Third, since the estimated sensitivity is larger than standard
predictions, any discovery would mean that our understanding of
supernovae was missing an effect more significant than neutrino
mixing.

The estimated sensitivity can also be approximately characterized by
the required flux in this interval, instead of the number of events.
Multiplying the flux, the number of deuterons, the cross section at 25
MeV, and the assumed 5-year exposure, we find $\phi \, (6 \times
10^{31}) \, (27 \times 10^{-42} \, {\rm cm}^{2}) (5 \, {\rm yrs}) \,
\simeq 1.6$ events, so that the required flux is 6 cm$^{-2}$ s$^{-1}$.
Ignoring the slight temperature dependence, we show this as a constant
in Fig.~\ref{fig:Boosts}.


\section{Electron Neutrinos from SN 1987A} 
\label{sec:1987A}

The neutrino flux from SN 1987A was observed with two water
\v{C}erenkov detectors, Kamiokande-II (Kam-II) and IMB
\cite{Hirata:1987hu,Hirata:1988ad,Bionta:1987qt,Bratton:1988ww}.  (And
likely also with the much smaller Baksan scintillator detector, which
had no directional sensitivity~\cite{Baksan}.)  The most detectable
component is the $\bar{\nu}_e$ flux, due to the largest cross section
being the inverse beta reaction $\bar{\nu}_e + p \rightarrow e^{+} +
n$, in which the outgoing positron angular distribution is nearly
isotropic~\cite{invbeta}.

Two features of the measured angular data remain mysterious.  One is
that the first event in Kam-II was directed forward.  The other is
that the majority of the 12 Kam-II events, and all but one of the 8
IMB events, were in the forward hemisphere.  Assuming that the yield
is dominated by inverse beta events, both features, and especially the
latter, are unlikely.  These features suggest a larger than expected
contribution from $\nu + e^- \rightarrow \nu + e^-$ scattering, in
which the electron closely follows the neutrino direction.  If so,
then the most likely possibility is that this was caused by an
enhanced flux of $\nu_e$; the $\bar{\nu}_e$ flux is constrained by the
inverse beta yield, and the other flavors have $\sim 6$ times smaller
cross sections on electrons.

Taking the measured data at face value, a few interpretations of the
noted features are possible.  One, that the $\nu_e$ flux in the early
neutronization burst phase was much larger than expected.  This could
explain the first forward event, but not the time-averaged angular
distribution.  Two, that the $\nu_e$ flux in the longer thermal phase
was much larger than expected.  This could help explain the
time-averaged angular distribution, and if the flux were large enough,
also make it likely that one of the $\nu_e + e^- \rightarrow \nu_e +
e^-$ events would happen to be the first event detected.  Third, that
there were statistical fluctuations, either upward fluctuations of the
small expected yields of $\nu_e + e^- \rightarrow \nu_e + e^-$
scattering, or forward fluctuations of the near-isotropic angular
distribution of $\bar{\nu}_e + p \rightarrow e^{+} + n$.  With the SN
1987A data alone, it not possible to decide among these
interpretations.  In the context of standard supernova models, the
first two interpretations are disfavored.  What we consider here is
that the required enhancements to the $\nu_e$ fluxes in the first two
interpretations would similarly increase the expectations for the DSNB
$\nu_e$ flux, and that this should be testable with SNO.

At Kam-II the first event was perfectly forward, with an angle $18 \pm
18$ degrees relative to the neutrino direction.  During the
neutronization phase, a prompt $\nu_e$ burst is created from electron
captures on newly-liberated protons, $e^{-} + p \rightarrow \nu_e +
n$~\cite{Bruenn,Sutaria,Rampp,Mezzacappa,Kachelriess}.  The fact that
the first event was {\it both} forward and first is suggestive of
its being caused by neutrino-electron scattering \cite{burstpro},
whatever the difficulties of explaining this with standard supernova
models \cite{Raffelt,Bruenn,Sato}.  (The relative timing of the Kam-II
and IMB detectors is unknown, and we are making the common assumption
that the first event in Kam-II was the first event overall.)  The
total energy released during this burst is expected to be about an
order of magnitude less than the $\nu_e$ energy released during the
thermal phase.  For a total $\nu_e$ burst energy of $4.5 \times
10^{51}$ erg~\cite{Bruenn}, the expected number of $\nu_e$ scatterings
at Kam-II is $\sim 0.01$ events, compared to $\sim 0.1$ for the
thermal phase.  These estimates neglect neutrino mixing, which can be
especially relevant for the burst phase, since the transformation of
$\nu_e$ into other flavors occurs without the reverse process
\cite{Walker,Ando:2004qe,Kachelriess}.

\begin{figure}
\includegraphics[width=3.25in]{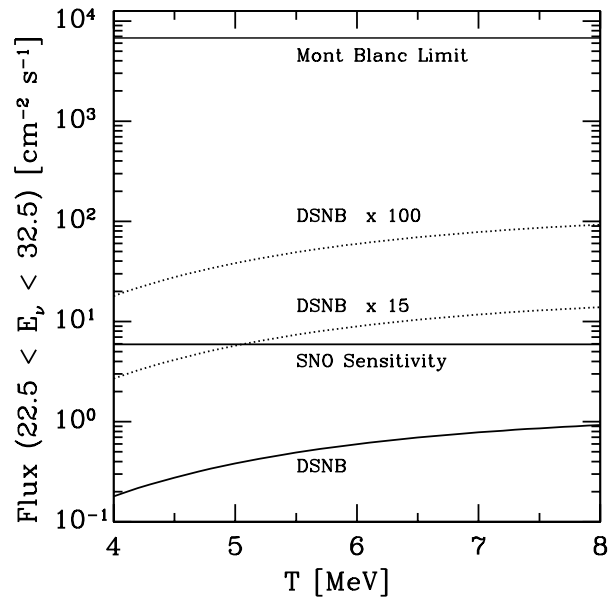}
\caption{\label{fig:Boosts}
The predicted DSNB flux as a function of the $\nu_e$ temperature.  The
approximate Mont Blanc limit and projected SNO sensitivity are shown
with solid lines.  The dotted lines approximately indicate the
enhanced DSNB $\nu_e$ fluxes expected if the apparent indications of
$\nu_e + e^- \rightarrow \nu_e + e^-$ events in the SN 1987A data are
interpreted as a probable outcome.  The upper dotted line corresponds
to assuming that the first Kam-II event was due to $\nu_e$ from the
neutronization burst phase.  The lower dotted line corresponds to
assuming instead that it arose from $\nu_e$ from the thermal phase
(and that in total $\sim 3$ of the events in Kam-II were due to
$\nu_e$).}
\end{figure}

An enhanced rate of $\nu_e + e^- \rightarrow \nu_e +e^-$ during the
thermal phase could help explain why the time-averaged angular
distributions were so forward~\cite{LoSecco,Kielczewska}.  A recent
reanalysis determines that a few of the Kam-II events, and at most one
of the IMB events, are consistent with elastic scattering
events~\cite{Costantini}.  With or without these events, the angular
distributions of Kam-II and IMB are not consistent with each other (or
with expectations), when one considers both the mean and the variance
of $\cos\theta$~\cite{invbeta,Costantini}.  However, as a caution
about the small-number statistics, note that had even a single
additional event been detected in a backward direction in IMB, then
the disagreements would have been less severe~\cite{invbeta}.

Some potential scenarios for increased $\nu_e$ emission have been
considered in Refs.~\cite{Arnett,Burrows,Imshennik,Costantini}.  A
boost in the $\nu_e$ yield relative to $\bar{\nu}_e$ can be
characterized by the factor
\begin{equation}
f_{\nu_e} = 
\frac{n}{N-n} \frac{\langle \sigma_{\bar{\nu}_e} \rangle / T_{\bar{\nu}_e}}
{\langle \sigma_{\nu_e} \rangle / T_{\nu_e}}.
\label{eq:boost} 
\end{equation}
Here $N$ is the total number of events observed at Kam-II or IMB ($12$
or $8$), $n$ is the number of $\nu_e + e^- \rightarrow \nu_e + e^-$
scatterings, and $\langle \sigma \rangle$ is the thermally averaged
cross-section, taking into account the thermal distribution and the
efficiency of the detector~\cite{Haxton:1987kc}.  In deriving
Eq.~(\ref{eq:boost}), we have assumed that there is equal total energy
emitted in each neutrino flavor.  Since during the burst phase the
total energy emitted in $\nu_e$ is approximately an order of magnitude
less than by $\nu_e$ in the thermal phase, $f_{\nu_e}$ increases by
$E_{\bar{\nu}_e}^{thermal}/E_{\nu_e}^{burst}$ when considering a
probable scenario of one event during the burst phase.

For an illustration for a typical value of $f_{\nu_e}$ for the thermal
phase we consider $n =3$ and $T_{\bar{\nu}_e} = 3$ MeV for Kam-II.  We
choose $n = 3$ since this is the smallest $n$ such that there is a
reasonable chance of having an electron scattering event be first, and
also so that the time-integrated angular distribution is affected at a
significant level.  A typical value is $f_{\nu_e} \sim 15$ over the
range of $T_{\nu_e}$.  In Fig.~\ref{fig:Boosts} we show what this
increase corresponds to in the DSNB flux, and how it compares to the
SNO sensitivity.  (If the $\nu_e$ is assumed to come from the
neutronization burst phase, the required boost is even larger, at
least $\sim 100$.)  At high assumed temperatures, the suggested
$\nu_e$ flux enhancements in the thermal phase would lead to too many
charged-current events on $^{16}$O in the SN 1987A data, which would
be an independent way to constrain these scenarios.  For example, for
standard fluxes, this yield would become a few events when $T = 8$
MeV; in addition, the angular distribution favors backward
angles~\cite{Haxton:1987kc}.  Also, for large temperatures and fluxes,
the number of events expected in the Homestake detector might become
too large \cite{Minakata:1987if}.

There also remains the possibility that the first Kam-II event
resulted from $\bar{\nu}_e + p \rightarrow e^{+} + n$, and just
happened to be emitted in the forward direction.  For example, there
is a $ \sim 5\%$ chance that any $\bar{\nu}_e$ event was emitted
within a given cone of half-angle 25 degrees, comparable to the
uncertainty in the Kam-II angular resolution.  To have the {\it first}
event be forward thus has a $\sim 5\%$ probability, and to explain the
time-averaged angular distribution would require additional
statistical fluctuations.  By searching for the DSNB $\nu_e$ flux, SNO
will be the first experiment able to revisit the electron-neutrino
component of the SN 1987A data and determine if the observed angular
distribution was indeed a probable outcome.


\section{Conclusions}

We have shown that SNO should have sensitivity to a Diffuse Supernova
Neutrino Background (DSNB) $\nu_e$ flux as low as $\simeq 6$ cm$^{-2}$
s$^{-1}$, which is about three orders of magnitude smaller than the
current limit from the Mont Blanc experiment~\cite{Aglietta}.  In
order to best avoid detector backgrounds, this is the flux just in the
energy range 22.5--32.5 MeV; our corresponding theoretical
predictions, given in Table~\ref{tab:table1}, are $\alt 1$ cm$^{-2}$
s$^{-1}$.  Our results depend on our estimates of the relevant solar
and atmospheric neutrino backgrounds, and how the latter might be
reduced.  To the extent possible, we based our estimates on measured
rates, especially the sub-\v{C}erenkov muon decay rate from SK.  Given
the very small projected statistics, a more precise estimate of the
sensitivity is not yet possible.  A full analysis by the SNO
Collaboration, using the measured data, is strongly encouraged.

Our estimate for the SNO sensitivity to DSNB $\nu_e$ may seem
surprisingly good, given that SNO is much smaller than SK.  However,
it can be confirmed by adjusting the existing SK DSNB $\bar{\nu}_e$
limit of 1.2 cm$^{-2}$ s$^{-1}$ for the difference in exposure,
ignoring the small differences in the considered energy ranges.  The
SK exposure was 22.5 kton for 4.1 years at about 50\% efficiency in
the relevant energy range, so about 45 kton-yr at full efficiency, or
about 9 times more than the assumed SNO exposure.  Since the SK limit
arises from a background-limited search, the scaling to 5 kton-yr in a
{\it light} water detector would worsen the limit by about $\sqrt{9} =
3$.  Taking into account the smaller signal cross section in a {\it
heavy} water detector (and assuming the same background rate) gives an
additional penalty of a factor $\simeq 2$, so that in the end the SNO
sensitivity should be about a factor of 6 worse than in SK, close to
what we deduced.  (Had SK been rate-limited, the estimated SNO
sensitivity would have instead been about 18 times worse.)  Note that
if the true SNO exposure is less than the 5 years of full efficiency
that we assumed, the flux sensitivity only changes by the square root
of the ratio of exposures.

An interesting related point is that once the KamLAND and SNO
detectors report on multiple kton-yr exposures, they should be able to
set DSNB $\bar{\nu}_e$ limits only a factor of a few to several weaker
than what SK has published so far.  This would be significant
principally because KamLAND and SNO should have lower backgrounds and
thresholds; the combination of these limits with the stronger limits
at higher energies in SK might lead to improved limits on the
supernova rates at higher redshift~\cite{Ando2002,Strigari}.  The KamLAND
and SNO limits so far~\cite{Eguchi,Aharmim2004} are significantly worse
than we are estimating here because they were based on the small
exposures of early data and considered narrower energy ranges.  In
addition, the SNO analysis was based on the pure D$_2$O phase, during
which the neutron detection efficiency was much lower than in the next
two phases.  The sensitivity improves linearly with the exposure
before the background-dominated regime is reached, after which it
improves only with the square root.  In terms of absolute sensitivity,
the signal requirements demand an exposure of at least several
kton-yr, beyond which we estimate that SNO will be
background-dominated.  (On a related issue, note that while the
present SK DSNB $\bar{\nu}_e$ sensitivity is background-limited, if SK
is enhanced by the addition of gadolinium, then the absolute
sensitivity will be greatly improved, and is projected to become
rate-limited~\cite{gadzooks}.)

Studying the DSNB with any $\nu_e$ channel other than $\nu_e + d$
requires a large cross section, large detector mass, and favorable kinematics.
In the future, this may be possible with large liquid argon detectors,
as long emphasized by Cline~\cite{Cline}.  For example, the estimated
DSNB $\nu_e$ sensitivity after a 15 kton-yr exposure of the ICARUS
detector is $\simeq 1.6$ cm$^{-2}$ s$^{-1}$~\cite{Cocco}, which would
be very interesting as a complement to the SK DSNB $\bar{\nu}_e$
sensitivity.  This estimate assumes that the only backgrounds in
ICARUS are those due to solar and atmospheric $\nu_e$.  Very likely,
other backgrounds will have to be considered as well.  For example, it
is assumed that the low-momentum (sub-\v{C}erenkov) muons, which are a
serious background in water-based detectors, could be completely
rejected in argon-based detectors~\cite{Cline,Cocco}.  In water-based
detectors, the muons with momenta below $\simeq 100$ MeV are
invisible.  In a liquid argon detector with an assumed kinetic energy
threshold of 5 MeV, muons with momenta less than $\simeq 30$ MeV would
also be invisible, and their decays would produce a relevant
background.  Also needed are studies of backgrounds induced by fast
neutrons entering the detector, as well as those arising from the
quenched light output of energetic charged particles.  Nevertheless,
this technique appears promising, and may have the potential to go
significantly beyond the sensitivity of SNO.

Besides waiting for another Milky Way supernova, in principle there is
another method to probe the $\nu_e$ emission.  The $\nu_e$ flux from
past Milky Way supernovae would have transformed terrestrial
molybdenum isotopes into unstable technetium isotopes (half-lives 2.6
and 4.2 million years), and the elemental ratios in deep molybdenum
deposits should reveal the received $\nu_e$ fluence over a comparable
time period~\cite{MoTc}.  To the best of our knowledge, the realistic
sensitivity of this technique is unknown, and no experiment is
planned.

In addition to improving the present DSNB $\nu_e$ sensitivity by about
three orders of magnitude, a full analysis by SNO can test whether the
large $\nu_e$ flux suggested by the SN 1987A data was probable, or was
simply a statistical fluctuation.  If the {\it first} event in
Kamiokande-II was really $\nu_e + e^- \rightarrow \nu_e + e^-$
scattering, and was a probable outcome, then the DSNB $\nu_e$ flux
should be greatly enhanced.  If this first event is assumed to be from
the neutronization phase, then the required enhancement is a factor at
least $\simeq 100$.  If this first event is assumed to be from the
thermal phase, the required enhancement is a factor at least $\simeq
15$; this would also mean that there should be at least a few other
$\nu_e$ events in the data, which would make the angular distributions
more forward, also as suggested by the data.  Both scenarios are
probably unrealistic, and our estimates are crude.  However, our point
is that at present, only SNO has the sensitivity to directly constrain
them, and for the first time be able to look back and shed light on
the mysteries still lingering from SN 1987A.


\begin{acknowledgments}
We thank David Cline, Raph Hix, Josh Klein, Terry Walker, and
especially Shin'ichiro Ando, Mark Vagins and Hasan Y\"{u}ksel for
helpful discussions.  J.~F.~B. is supported by The Ohio State
University, and L.~E.~S. by Department of Energy grant
DE-FG02-91ER40690.
\end{acknowledgments}



\end{document}